\documentclass[10pt]{article}
\usepackage{graphicx}
\usepackage{amsmath}
\usepackage{amssymb}
\usepackage{caption2}
\begin{document}
\bibliographystyle{prsty}
\begin{center}
{\large {\bf \sc{  Analysis of  the  $Z(4430)$ as the first radial excitation of the $Z_c(3900)$ }}} \\[2mm]
Zhi-Gang  Wang \footnote{E-mail: zgwang@aliyun.com.  }     \\
 Department of Physics, North China Electric Power University, Baoding 071003, P. R. China
\end{center}

\begin{abstract}
In this article, we  take the $Z_c(3900)$ and $Z(4430)$  as the ground state and the first radial excited state of the axial-vector  tetraquark states with $J^{PC}=1^{+-}$, respectively,
and study their masses and pole residues with the QCD sum rules by  calculating the contributions of the vacuum condensates up to dimension-10 in a
consistent way in the operator product expansion.  The numerical result favors assigning  the $Z_c(3900)$ and $Z(4430)$
 as the ground state and first radial excited state of the axial-vector  tetraquark states, respectively.
\end{abstract}

 PACS number: 12.39.Mk, 12.38.Lg

Key words: Tetraquark  state, QCD sum rules

\section{Introduction}

In 2007, the  Belle collaboration   observed a distinct peak   in the $\pi^{\pm} \psi^\prime$ invariant mass distribution  in the $B\to K \pi^{\pm} \psi^{\prime}$ decays
with the statistical significance of  $6.5\sigma$,
 the mass and width are $ M=\left(4433\pm4\pm2\right)\,\rm{ MeV}$ and $\Gamma = \left(45^{+18}_{-13} {}^{+30}_{-13}\right)\,\rm {MeV}$, respectively \cite{Belle-2007}.
In 2009, the  Belle collaboration observed a signal for the decay $Z(4430)^+ \to \pi^+ \psi^\prime$ with a mass $M= \left(4443^{+15}_{-12}{}^{+19}_{-13}\right)\,\rm{MeV}$ and a width
$\Gamma= \left(107^{+86}_{-43}{}^{+74}_{-56}\right)\,\rm{MeV}$ with a significance of $6.4\sigma$  from a Dalitz plot analysis of the decays  $B \to K \pi^+ \psi^\prime$ \cite{Belle-2009}.
In 2013, the Belle collaboration performed a full amplitude analysis of the $B^0 \to \psi^{\prime} K^+ \pi^-$ decays
 to constrain the spin and parity of the $Z(4430)^-$, and observed the $J^P=1^+$ hypothesis is favored over the $0^-$, $1^-$, $2^-$ and $2^+$ hypotheses
 at the levels of $3.4 \sigma$, $3.7 \sigma$, $4.7 \sigma$ and $5.1 \sigma$, respectively \cite{Belle-2013}.
 Recently, the LHCb collaboration analyzed the $B^0\to\psi'\pi^-K^+$ decays    by performing a four-dimensional fit of the decay amplitude
 using $pp$ collision data corresponding to $\rm 3 fb^{-1}$ collected with the LHCb detector, and  provided the first independent confirmation of
the existence of the $Z(4430)^-$ resonance
and established its spin-parity to be $1^+$.
The measured mass and width are $M=\left(4475\pm7\,{_{-25}^{+15}}\right)\,\rm {MeV}$ and
  $\Gamma=\left(172\pm13\,{_{-34}^{+37}}\right)\,\rm {MeV}$, respectively \cite{LHCb-1404}.
There have been several tentative assignments of the $Z(4430)$, such as
the   threshold effect   \cite{Z4430-threshold},  molecular state  \cite{Z4430-molecule},  tetraquark state \cite{Z4430-Tetraquark,Z4430-Maiani,Z4430-zgwang},
baryonium \cite{Z4430-CFQiao},
   hadro-charmonium state \cite{Z4430-Voloshin},   etc.

In 2013, the BESIII collaboration studied  the process  $e^+e^- \to \pi^+\pi^-J/\psi$ and observed a structure $Z_c(3900)$ in the $\pi^\pm J/\psi$ mass spectrum with a mass of $(3899.0\pm 3.6\pm 4.9)\,\rm{ MeV}$
and a width of $(46\pm 10\pm 20) \,\rm{MeV}$ \cite{BES3900}. Then the structure $Z_c(3900)$ was confirmed by the Belle and CLEO collaborations \cite{Belle3900,CLEO3900}.
 R. Faccini et al tentatively identify the $Z_c(3900)$ as the negative charge conjugation partner of the $X(3872)$ \cite{Maiani1303}, other assignments,
 such as molecular state \cite{Molecular3900}, tetraquark state \cite{Tetraquark3900}, hadro-charmonium \cite{hadro-charmonium-3900}, rescattering effect \cite{FSI3900},
 are also suggested.  In Ref.\cite{Z4430-1405}, L. Maiani et al take the  $Z(4430)$ as the first radial excitation of the $Z_c(3900)$ according to the
analogous decays,
\begin{eqnarray}
Z_c(3900)^\pm&\to&J/\psi\pi^\pm\, , \nonumber \\
Z(4430)^\pm&\to&\psi^\prime\pi^\pm\, .
\end{eqnarray}
The mass differences are $M_{Z(4430)}-M_{Z_c(3900)}=576\,\rm{MeV}$ and $M_{\psi^\prime}-M_{J/\psi}=589\,\rm{MeV}$, so it is natural to take
 the  $Z(4430)$ as the first radial
excitation of the $Z_c(3900)$ \cite{Nielsen-1401}.

The  QCD sum rules is a powerful nonperturbative theoretical tool in studying the
ground state hadrons \cite{SVZ79,Reinders85}.
In Refs.\cite{WangHuangTao,Wang1311}, we focus on the scenario of tetraquark states, calculate the  vacuum condensates up to dimension-10  in
the operator product expansion, study the diquark-antidiquark type scalar, vector, axial-vector, tensor hidden charmed tetraquark states and
axial-vector hidden bottom tetraquark states systematically  with the QCD sum rules, and make reasonable  assignments of the $X(3872)$,
$Z_c(3900)$, $Z_c(3885)$, $Z_c(4020)$, $Z_c(4025)$, $Z(4050)$, $Z(4250)$, $Y(4360)$, $Y(4630)$, $Y(4660)$, $Z_b(10610)$  and $Z_b(10650)$.
In Ref.\cite{WangHuangTao1312}, we focus on the scenario of  molecular states,
 calculate the  vacuum condensates up to dimension-10  in
the operator product expansion, study the scalar, axial-vector and tensor  hadronic molecular states with the QCD sum rules, and make tentative assignments  of the
$X(3872)$, $Z_c(3900)$,  $Y(3940)$, $Y(4140)$, $Z_c(4020)$, $Z_c(4025)$, $Z_b(10610)$ and $Z_b(10650)$.
  In Refs.\cite{WangHuangTao,Wang1311,WangHuangTao1312}, we   explore the energy scale dependence of the hidden charmed (bottom) tetraquark states and molecular states in details for the first time, and suggest a  formula
\begin{eqnarray}
\mu&=&\sqrt{M^2_{X/Y/Z}-(2{\mathbb{M}}_Q)^2} \, ,
 \end{eqnarray}
 with the effective masses ${\mathbb{M}}_Q$ to determine  the energy scales of the  QCD spectral densities in the QCD sum rules, which works very well.

In this article, we extend our previous work on the $X(3872)$, $Z_c(3900)$, $Z_c(3885)$ \cite{WangHuangTao},  focus on the scenario of tetraquark states,
take the $Z_c(3900)$ and $Z(4430)$  as the ground state and first radial excited state of the axial-vector  tetraquark states with the symbolic quark structure
$[cu]_{S=1}[\bar{c}\bar{d}]_{S=0}-[cu]_{S=0}[\bar{c}\bar{d}]_{S=1}$,
and study them with the QCD sum rules.

The article is arranged as follows:  we derive the QCD sum rules for
the masses and pole residues of  the axial-vector tetraquark states $Z_c(3900)$ and  $Z(4430)$ in section 2; in section 3,
we present the numerical results and discussions; section 4 is reserved for our conclusion.

\section{QCD sum rules for  the   $J^{PC}=1^{+-}$ tetraquark states }
In the following, we write down  the two-point correlation function $\Pi_{\mu\nu}(p)$  in the QCD sum rules,
\begin{eqnarray}
\Pi_{\mu\nu}(p)&=&i\int d^4x e^{ip \cdot x} \langle0|T\left\{J_\mu(x)J_\nu^{\dagger}(0)\right\}|0\rangle \, , \\
J_\mu(x)&=&\frac{\epsilon^{ijk}\epsilon^{imn}}{\sqrt{2}}\left\{u^j(x)C\gamma_5c^k(x) \bar{d}^m(x)\gamma_\mu C \bar{c}^n(x)-u^j(x)C\gamma_\mu c^k(x)\bar{d}^m(x)\gamma_5C \bar{c}^n(x) \right\} \, ,
\end{eqnarray}
 the $i$, $j$, $k$, $m$, $n$ are color indexes, the $C$ is the charge conjugation matrix. We choose  the   current $J_\mu(x)$ to interpolate the
  $J^{PC}=1^{+-}$ diquark-antidiquark type tetraquark states  $Z_c(3900)$ and $Z(4430)$.  Under charge conjugation transform $\widehat{C}$,
  the current $J_\mu(x)$ has the property,
\begin{eqnarray}
\widehat{C}J_{\mu}(x)\widehat{C}^{-1}&=&- J_\mu(x)\mid_{u \leftrightarrow d} \, ,
\end{eqnarray}
which originates  from the charge conjugation  properties of the scalar and axial-vector diquark states,
\begin{eqnarray}
\widehat{C}\left[\epsilon^{ijk}q^j C\gamma_5 c^k\right]\widehat{C}^{-1}&=&\epsilon^{ijk}\bar{q}^j \gamma_5 C \bar{c}^k \, , \nonumber\\
\widehat{C}\left[\epsilon^{ijk}q^j C\gamma_\mu c^k\right]\widehat{C}^{-1}&=&\epsilon^{ijk}\bar{q}^j \gamma_\mu C \bar{c}^k \, .
\end{eqnarray}

We can insert  a complete set of intermediate hadronic states with
the same quantum numbers as the current operator $J_\mu(x)$ into the
correlation function $\Pi_{\mu\nu}(p)$  to obtain the hadronic representation
\cite{SVZ79,Reinders85}. After isolating the ground state
and the first radial excited state contributions from the pole terms, which are supposed to be the tetraquark states   $Z_c(3900)$  and $Z(4430)$, we get the following results,
\begin{eqnarray}
\Pi_{\mu\nu}(p)&=&\left[\frac{\lambda_{Z_c(3900)}^2}{M_{Z_c(3900)}^2-p^2}+\frac{\lambda_{Z(4430)}^2}{M_{Z(4430)}^2-p^2}\right]\left(-g_{\mu\nu} +\frac{p_\mu p_\nu}{p^2}\right) +\cdots \, \, ,\\
&=&\Pi(p^2)\left(-g_{\mu\nu} +\frac{p_\mu p_\nu}{p^2}\right) +\cdots \, ,
\end{eqnarray}
where the pole residues   $\lambda_{Z}$ are defined by
\begin{eqnarray}
 \langle 0|J_\mu(0)|Z(p)\rangle=\lambda_{Z} \, \varepsilon_\mu \, ,
\end{eqnarray}
the $\varepsilon_\mu$ are the polarization vectors of the axial-vector mesons $Z_c(3900)$ and $Z(4430)$. The current $J_\mu(x)$ has the $J^{PC}=1^{+-}$,
the $Z_c(3900)$ and $Z(4430)$ also have the $J^{PC}=1^{+-}$ according to the decays $Z_c(3900)^\pm\to J/\psi\pi^\pm$
and $Z(4430)^\pm\to\psi^\prime\pi^\pm$. The final states $J/\psi \pi^\pm$ and $\psi^\prime \pi^\pm$ indicate that the $Z_c(3900)$ and $Z(4430)$ must have some $c\bar{c}u\bar{d}$ or $c\bar{c}d\bar{u}$ components at the quark level. The current $J_\mu(x)$ couples  potentially to the $Z_c(3900)$ and $Z(4430)$. On the other hand, the   current $J_\mu(x)$  has non-vanishing couplings  with the scattering states  $D
D^\ast$, $J/\psi \pi$, $J/\psi \rho$, $\cdots$  \cite{PDG}. The coupling to the intermediate  scattering states  $D
D^\ast$, $J/\psi \pi$, $J/\psi \rho$, $\cdots$ modifies the hadronic states $Z_c(3900)$ and $Z(4430)$ through self-energy corrections \cite{WangHuangTao}.
The renormalized self-energies  contribute  a finite imaginary part to modify the dispersion relation \cite{WangHuangTao},
\begin{eqnarray}
\Pi(p^2) &=&-\frac{\lambda_{Z_c(3900)}^{2}}{ p^2-M_{Z_c(3900)}^2+i\sqrt{p^2}\Gamma_{Z_c(3900)}(p^2)}-\frac{\lambda_{Z(4430)}^{2}}{ p^2-M_{Z(4430)}^2+i\sqrt{p^2}\Gamma_{Z(4430)}(p^2)}+\cdots \, , \nonumber\\
 \end{eqnarray}
where the physical  widths $\Gamma_{Z_c(3900)}\left(M_{Z_c(3900)}^2\right)=\left(46 \pm 10 \pm 20\right)\, \rm{MeV}$ and  $\Gamma_{Z(4430)}\left(M_{Z(4430)}^2\right)=\left(172\pm13\,{_{-34}^{+37}}\right)\,\rm {MeV}$ are not very large,
 the zero width approximation in  the hadronic spectral densities works \cite{WangIJTP}.

We carry out the
operator product expansion to the vacuum condensates up to dimension-10 and
take the assumption of vacuum saturation for the  higher dimension vacuum condensates.
The condensates $\langle \frac{\alpha_s}{\pi}GG\rangle$, $\langle \bar{q}q\rangle\langle \frac{\alpha_s}{\pi}GG\rangle$,
$\langle \bar{q}q\rangle^2\langle \frac{\alpha_s}{\pi}GG\rangle$, $\langle \bar{q} g_s \sigma Gq\rangle^2$ and $g_s^2\langle \bar{q}q\rangle^2$ are the vacuum expectations
of the operators of the order
$\mathcal{O}(\alpha_s)$.
 The condensates $\langle g_s^3 GGG\rangle$, $\langle \frac{\alpha_s GG}{\pi}\rangle^2$,
 $\langle \frac{\alpha_s GG}{\pi}\rangle\langle \bar{q} g_s \sigma Gq\rangle$ have the dimensions 6, 8, 9 respectively,  but they are   the vacuum expectations
of the operators of the order    $\mathcal{O}( \alpha_s^{3/2})$, $\mathcal{O}(\alpha_s^2)$, $\mathcal{O}( \alpha_s^{3/2})$ respectively, and discarded.  We take
the truncations $n\leq 10$ and $k\leq 1$ in a consistent way,
the operators of the orders $\mathcal{O}( \alpha_s^{k})$ with $k> 1$ are  discarded. Furthermore,  the values of the  condensates $\langle g_s^3 GGG\rangle$, $\langle \frac{\alpha_s GG}{\pi}\rangle^2$,
 $\langle \frac{\alpha_s GG}{\pi}\rangle\langle \bar{q} g_s \sigma Gq\rangle$   are very small, and they can be  neglected safely.
 For the technical details, one can consult Ref.\cite{WangHuangTao}.

 Once the QCD spectral densities  are obtained,  we can take the
quark-hadron duality below the continuum threshold $s_0$ and perform Borel transform  with respect to
the variable $P^2=-p^2$ to obtain  the following QCD sum rule:
\begin{eqnarray}
\Pi(T^2)&=&\lambda_{Z_c(3900)}^2\, \exp\left(-\frac{M_{Z_c(3900)}^2}{T^2}\right)+\lambda_{Z(4430)}^2\,
 \exp\left(-\frac{M_{Z(4430)}^2}{T^2}\right)\, ,\nonumber\\
 &=& \int_{4m_c^2}^{s_0} ds\, \rho(s) \, \exp\left(-\frac{s}{T^2}\right) \, . 
\end{eqnarray}
One can consult Ref.\cite{WangHuangTao} for the explicit expression of the QCD spectral density $\rho(s)$.

In Ref.\cite{Baxi-G}, M. S. Maior de Sousa and R. Rodrigues da Silva introduce a new approach  to calculate the masses and decay constants of the ground state and first radial excited state with the QCD sum rules. Furthermore, they study the masses and decay constants of the $\rho({\rm 1S,2S})$, $\psi({\rm 1S,2S})$, $\Upsilon({\rm 1S,2S})$     as an application, and observe that the ground state masses are smaller than the experimental values, which is explained as a shortcoming of this approach. In this article, we apply the approach  to study the heavy tetraquark systems, and resort to the energy scale formula
\begin{eqnarray}
\mu&=&\sqrt{M^2_{X/Y/Z}-(2{\mathbb{M}}_c)^2}\, ,
\end{eqnarray}
where the effective $c$-quark mass  ${\mathbb{M}}_c=1.8\,\rm{GeV}$ \cite{WangHuangTao,Wang1311,WangHuangTao1312},
to overcome the shortcoming \cite{Baxi-G}.

In the following, we will repeat the steps and write the expressions in a compact form.
Now let us introduce the notations $\tau=\frac{1}{T^2}$, $D^n=\left( -\frac{d}{d\tau}\right)^n$, and use the subscripts $1$ and $2$ to denote the ground state ($Z_c(3900)$) and the first excited state $Z(4430)$, respectively, then the QCD sum rule can be written as
\begin{eqnarray}
\lambda_1^2\exp\left(-\tau M_1^2 \right)+\lambda_2^2\exp\left(-\tau M_2^2 \right)&=&\Pi_{QCD}(\tau) \, .
\end{eqnarray}
We differentiate  the QCD sum rule with respect to $\tau$ to obtain
\begin{eqnarray}
\lambda_1^2M_1^2\exp\left(-\tau M_1^2 \right)+\lambda_2^2M_2^2\exp\left(-\tau M_2^2 \right)&=&D\Pi_{QCD}(\tau) \, .
\end{eqnarray}
Now we have two equations, it is easy to obtain the sum rules,
\begin{eqnarray}
\lambda_i^2\exp\left(-\tau M_i^2 \right)&=&\frac{\left(D-M_j^2\right)\Pi_{QCD}(\tau)}{M_i^2-M_j^2} \, ,
\end{eqnarray}
where $i \neq j$.
Again we differentiate  above  QCD sum rules with respect to $\tau$ to obtain
\begin{eqnarray}
M_i^2&=&\frac{\left(D^2-M_j^2D\right)\Pi_{QCD}(\tau)}{\left(D-M_j^2\right)\Pi_{QCD}(\tau)} \, , \nonumber\\
M_i^4&=&\frac{\left(D^3-M_j^2D^2\right)\Pi_{QCD}(\tau)}{\left(D-M_j^2\right)\Pi_{QCD}(\tau)}\, .
\end{eqnarray}
 The squared masses $M_i^2$ satisfy the following equation,
\begin{eqnarray}
M_i^4-b M_i^2+c&=&0\, ,
\end{eqnarray}
where
\begin{eqnarray}
b&=&\frac{D^3\otimes D^0-D^2\otimes D}{D^2\otimes D^0-D\otimes D}\, , \nonumber\\
c&=&\frac{D^3\otimes D-D^2\otimes D^2}{D^2\otimes D^0-D\otimes D}\, , \nonumber\\
D^j \otimes D^k&=&D^j\Pi_{QCD}(\tau) \,  D^k\Pi_{QCD}(\tau)\, ,
\end{eqnarray}
$i=1,2$, $j,k=0,1,2,3$.
The solutions are
\begin{eqnarray}
M_1^2=\frac{b-\sqrt{b^2-4c} }{2} \, ,\nonumber\\
M_2^2=\frac{b+\sqrt{b^2-4c} }{2} \, .
\end{eqnarray}
\section{Numerical results and discussions}
The input parameters are taken to be the standard values $\langle
\bar{q}q \rangle=-(0.24\pm 0.01\, \rm{GeV})^3$,   $\langle
\bar{q}g_s\sigma G q \rangle=m_0^2\langle \bar{q}q \rangle$,
$m_0^2=(0.8 \pm 0.1)\,\rm{GeV}^2$, $\langle \frac{\alpha_s
GG}{\pi}\rangle=(0.33\,\rm{GeV})^4 $    at the energy scale  $\mu=1\, \rm{GeV}$
\cite{SVZ79,Reinders85,Ioffe2005,ColangeloReview}.
The quark condensate and mixed quark condensate evolve with the   renormalization group equation, $\langle\bar{q}q \rangle(\mu^2)=\langle\bar{q}q \rangle(Q^2)\left[\frac{\alpha_{s}(Q)}{\alpha_{s}(\mu)}\right]^{\frac{4}{9}}$ and $\langle\bar{q}g_s \sigma Gq \rangle(\mu^2)=\langle\bar{q}g_s \sigma Gq \rangle(Q^2)\left[\frac{\alpha_{s}(Q)}{\alpha_{s}(\mu)}\right]^{\frac{2}{27}}$.

In the article, we take the $\overline{MS}$ mass $m_{c}(m_c^2)=(1.275\pm0.025)\,\rm{GeV}$
 from the Particle Data Group \cite{PDG}, and take into account
the energy-scale dependence of  the $\overline{MS}$ mass from the renormalization group equation,
\begin{eqnarray}
m_c(\mu^2)&=&m_c(m_c^2)\left[\frac{\alpha_{s}(\mu)}{\alpha_{s}(m_c)}\right]^{\frac{12}{25}} \, ,\nonumber\\
\alpha_s(\mu)&=&\frac{1}{b_0t}\left[1-\frac{b_1}{b_0^2}\frac{\log t}{t} +\frac{b_1^2(\log^2{t}-\log{t}-1)+b_0b_2}{b_0^4t^2}\right]\, ,
\end{eqnarray}
  where $t=\log \frac{\mu^2}{\Lambda^2}$, $b_0=\frac{33-2n_f}{12\pi}$, $b_1=\frac{153-19n_f}{24\pi^2}$, $b_2=\frac{2857-\frac{5033}{9}n_f+\frac{325}{27}n_f^2}{128\pi^3}$,  $\Lambda=213\,\rm{MeV}$, $296\,\rm{MeV}$  and  $339\,\rm{MeV}$ for the flavors  $n_f=5$, $4$ and $3$, respectively  \cite{PDG}.

  The mass and width from the LHCb collaboration   are $M_{Z(4430)}=\left(4475\pm7\,{_{-25}^{+15}}\right)\,\rm {MeV}$ and
  $\Gamma_{Z(4430)}=\left(172\pm13\,{_{-34}^{+37}}\right)\,\rm {MeV}$, respectively \cite{LHCb-1404}, we can take the
continuum  threshold parameter as $\sqrt{s_0}=(4.7-4.9)\,\rm{GeV}$  tentatively
 to avoid the contaminations from  the higher  resonances and continuum states, here we have assumed that the energy gap between the first radial excited state and
 the second radial excited state is about $(0.3\pm0.1)\,\rm{GeV}$, which is smaller than the energy gap $(0.5\pm0.1)\,\rm{GeV}$ between the ground state and
 the first radial excited state. If we take the Borel parameter as $T^2=(2.7-3.3)\,\rm{GeV}^2$, the pole contribution is $(55-80)\%$ ($(64-86)\%$) at the typical
  energy scale $\mu=1.5\,\rm{GeV}$ ($\mu=2.7\,\rm{GeV}$).
 In Ref.\cite{WangHuangTao}, we take the parameters  $\sqrt{s_0}=(4.3-4.5)\,\rm{GeV}$, $T^2=(2.2-2.8)\,\rm{GeV}^2$ and $\mu=1.5\,\rm{GeV}$ to
 study the ground state $Z_c(3900)$, the two criteria (pole dominance and convergence of the operator product
expansion) of the QCD sum rules are fully satisfied. In the present case, if we take the  parameters  $\sqrt{s_0}=(4.7-4.9)\,\rm{GeV}$, $T^2=(2.4-3.8)\,\rm{GeV}^2$
and $\mu=1.5\,\rm{GeV}$, the two criteria   of the QCD sum rules are also satisfied.

Firstly, we choose the continuum  threshold parameter as $\sqrt{s_0}=(4.7-4.9)\,\rm{GeV}$ and the Borel parameter as $T^2=(2.4-3.8)\,\rm{GeV}^2$, take the masses $M_{Z_c(3900)}=3899\,\rm{MeV}$ and $M_{Z(4430)}=4475\,\rm{MeV}$ as input parameters,   fit the pole residues $\lambda_{Z_c(3900)}$ and $\lambda_{Z(4430)}$ as free parameters with the
MINUIT, and obtain the results,
\begin{eqnarray}
\lambda_{Z_c(3900)}&=&(1.9977  \pm 0.0856)\times 10^{-2}\,\rm{GeV}^5 \, , \nonumber\\
\lambda_{Z(4430)}&=&(3.6186  \pm 0.2248)\times 10^{-2}\,\rm{GeV}^5 \,   ,
\end{eqnarray}
at the energy scale $\mu=1.5\,\rm{GeV}$ and
\begin{eqnarray}
\lambda_{Z_c(3900)}&=&(3.5125  \pm 0.4098)\times 10^{-2}\,\rm{GeV}^5 \, , \nonumber\\
\lambda_{Z(4430)}&=&(3.3554  \pm 1.9965)\times 10^{-2}\,\rm{GeV}^5 \,   ,
\end{eqnarray}
at the energy scale $\mu=2.7\,\rm{GeV}$.

In Ref.\cite{WangHuangTao}, we obtain the mass and pole residue of the $Z_c(3900)$ with the single pole QCD sum rules,
\begin{eqnarray}
M_{Z_c(3900)}&=&3.91^{+0.11}_{-0.09}\,\rm{GeV} \, ,  \nonumber\\
\lambda_{Z_c(3900)}&=&2.20^{+0.36}_{-0.29}\times 10^{-2}\,\rm{GeV}^5 \,   .
\end{eqnarray}
The value of the pole residue $\lambda_{Z_c(3900)}$ obtained in the present work at the energy scale $\mu=1.5\,\rm{GeV}$ is compatible with that of Ref.\cite{WangHuangTao}.
In Fig.1, we plot the central values of the Borel transformed correlation function $\Pi(T^2)$ at both the QCD side and the hadron side at the energy scale $\mu=1.5\,\rm{GeV}$. From the figure, we can see that the two curves
coincide, the fitting is excellent, on the other hand, the corresponding two curves also coincide at the energy scale $\mu=2.7\,\rm{GeV}$. At the two typical energy scales $\mu=1.5\,\rm{GeV}$ and $2.7\,\rm{GeV}$, we can take the masses of the $Z_c(3900)$ and $Z(4430)$ as basic input parameters, and choose
  suitable pole residues to reproduce the Borel transformed correlation function $\Pi(T^2)$ at the QCD side.
   The QCD sum rules favor assigning the  $Z(4430)$ as the first radial excitation of the $Z_c(3900)$ with the $J^{PC}=1^{+-}$.
\begin{figure}
 \centering
 \includegraphics[totalheight=8cm,width=12cm]{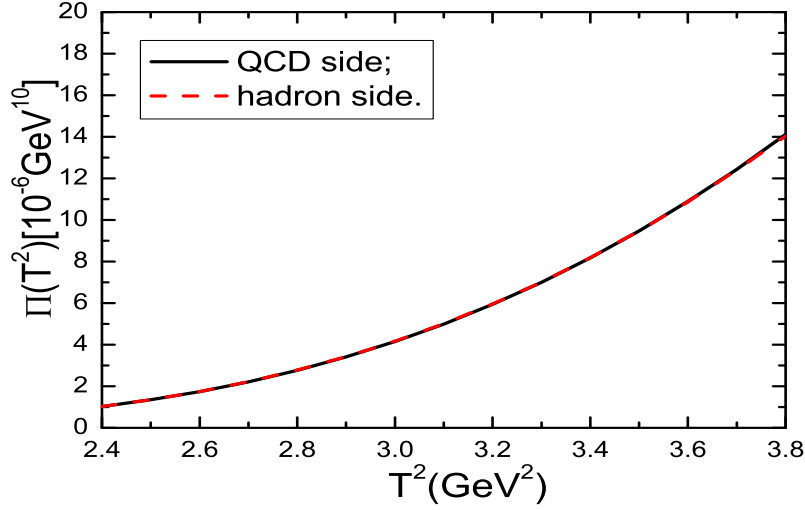}
    \caption{The central values of the Borel transformed correlation function. }
\end{figure}

In Refs.\cite{WangHuangTao,Wang1311,WangHuangTao1312}, we calculate the  vacuum condensates up to dimension-10  in
the operator product expansion, study the  hidden charmed (bottom) tetraquark states and
 molecular states systematically  with the QCD sum rules,  and  explore the energy scale dependence of the hidden charmed (bottom) tetraquark states and molecular states in details for the first time, and suggest a  formula
\begin{eqnarray}
\mu&=&\sqrt{M^2_{X/Y/Z}-(2{\mathbb{M}}_Q)^2} \, ,
 \end{eqnarray}
  to determine  the energy scales of the  QCD spectral densities.
 In the present case, if we resort to the formulaes in Eqs.(15-19) to study the masses and pole residues of the $Z_c(3900)$ and $Z(4430)$ as the ground state and the
  first radial excited state of the $J^{PC}=1^{+-}$ tetraquark states, respectively, the optimal energy scales are $\mu=1.5\,\rm{GeV}$ and $2.7\,\rm{GeV}$ for the QCD sum rules of the $Z_c(3900)$ and $Z(4430)$, respectively, the shortcoming in Ref.\cite{Baxi-G} is overcome.
  At the energy scale $\mu=1.5\,\rm{GeV}$ ($2.7\,\rm{GeV}$), we can obtain the physical value $M_{Z_c(3900)}$ ($M_{Z(4430)}$), the associate value $M_{(4430)}$ ($M_{Z_c(3900)}$) from the coupled Eqs.(18-19) is not necessary the physical value, and is discarded.

 Now we take into account the uncertainties and obtain the values of the masses and pole residues of the $Z_c(3900)$ and $Z(4430)$,
\begin{eqnarray}
M_{Z_c(3900)}&=&3.91^{+0.21}_{-0.17}\,\rm{GeV} \, ,  \,\,\, {\rm Experimental\,\, value} \,\,\,\,3899.0\pm 3.6\pm 4.9\,\rm{ MeV} \, \cite{BES3900}\,   , \nonumber\\
M_{Z(4430)}&=&4.70^{+0.98}_{-0.17}\,\rm{GeV} \, ,  \nonumber\\
\lambda_{Z_c(3900)}&=&2.23^{+1.02}_{-0.58}\times 10^{-2}\,\rm{GeV}^5 \, , \nonumber\\
\lambda_{Z(4430)}&=&4.19^{+3.83}_{-0.76}\times 10^{-2}\,\rm{GeV}^5 \,   ,
\end{eqnarray}
at the energy scale $\mu=1.5\,\rm{GeV}$ and
\begin{eqnarray}
M_{Z_c(3900)}&=&3.58^{+0.16}_{-0.11}\,\rm{GeV} \,   , \nonumber\\
M_{Z(4430)}&=&4.51^{+0.17}_{-0.09}\,\rm{GeV} \, ,  \,\,\, {\rm Experimental\,\, value} \,\,\,\,4475\pm7\,{_{-25}^{+15}}\,\rm {MeV}\, \cite{LHCb-1404}\,   , \nonumber\\
\lambda_{Z_c(3900)}&=&1.95^{+0.61}_{-0.26}\times 10^{-2}\,\rm{GeV}^5 \, , \nonumber\\
\lambda_{Z(4430)}&=&5.75^{+0.98}_{-0.78}\times 10^{-2}\,\rm{GeV}^5 \,   ,
\end{eqnarray}
at the energy scale $\mu=2.7\,\rm{GeV}$.
Then we take the central values of the masses and pole residues, and obtain the corresponding  pole contributions,
\begin{eqnarray}
{\rm pole}_{Z_c(3900)}&=&(38-62)\% \, , \nonumber \\
{\rm pole}_{Z(4430)}&=&(17-18)\% \, ,
\end{eqnarray}
at the energy scale $\mu=1.5\,\rm{GeV}$ and
\begin{eqnarray}
{\rm pole}_{Z_c(3900)}&=&(34-56)\% \, , \nonumber \\
{\rm pole}_{Z(4430)}&=&30\% \, ,
\end{eqnarray}
at the energy scale $\mu=2.7\,\rm{GeV}$. The pole contribution of the $Z_c(3900)$ ($Z(4430)$) at the energy scale $\mu=1.5\,\rm{GeV}$ ($2.7\,\rm{GeV}$) is a larger   than that at the energy scale $\mu=2.7\,\rm{GeV}$ ($1.5\,\rm{GeV}$), we prefer to extract the mass and pole residue of the $Z_c(3900)$ ($Z(4430)$) at the energy scale $\mu=1.5\,\rm{GeV}$ ($2.7\,\rm{GeV}$) and discard the ones at  the energy scale $\mu=2.7\,\rm{GeV}$ ($1.5\,\rm{GeV}$), and refer to the values  $M_{Z_c(3900)}=3.91^{+0.21}_{-0.17}\,\rm{GeV}$, $M_{Z(4430)}=4.51^{+0.17}_{-0.09}\,\rm{GeV} $, $\lambda_{Z_c(3900)}=2.23^{+1.02}_{-0.58}\times 10^{-2}\,\rm{GeV}^5$, $\lambda_{Z(4430)}=5.75^{+0.98}_{-0.78}\times 10^{-2}\,\rm{GeV}^5$ as the physical values, which are shown explicitly in Figs.2-3. The predicted masses $M_{Z_c(3900)}=3.91^{+0.21}_{-0.17}\,\rm{GeV}$ and  $M_{Z(4430)}=4.51^{+0.17}_{-0.09}\,\rm{GeV} $ satisfy the energy scale formula in Eq.(24).

The predicted masses $M_{Z_c(3900)}=3.91^{+0.21}_{-0.17}\,\rm{GeV}$ and  $M_{Z(4430)}=4.51^{+0.17}_{-0.09}\,\rm{GeV} $ are in excellent agreement with the experimental data, the present calculations favor assigning the $Z(4430)$ as  the first radial excited state of the $Z_c(3900)$. At the energy scale  $\mu=1.5\,\rm{GeV}$, the values of the pole residue $\lambda_{Z_c(3900)}$ from the numerical fitting, the single-pole QCD sum rules \cite{WangHuangTao} and the double-pole QCD sum rules are consistent with each other. At the energy scale  $\mu=2.7\,\rm{GeV}$, the values of the pole residue $\lambda_{Z(4430)}$ from the numerical fitting and the double-pole QCD sum rules are not consistent, as the pole residue $\lambda_{Z(4430)}$ is sensitive to the mass $M_{Z(4430)}$.

The parameters $M_{Z_c(3900)}$, $M_{Z(4430)}$, $\lambda_{Z_c(3900)}$, $\lambda_{Z(4430)}$ are not independent, they correlate with each other. For example, at the neighborhood of the values $M_{Z_c(3900)}=3.899\,\rm{GeV}$, $M_{Z(4430)}=4.475\,\rm{GeV}$, $\lambda_{Z_c(3900)}=1.9977  \times 10^{-2}\,\rm{GeV}^5$,
$\lambda_{Z(4430)}=3.6186  \times 10^{-2}\,\rm{GeV}^5$,
we can obtain the relations,
\begin{eqnarray}
M_{Z_c(3900)}\uparrow  &\longmapsto&  M_{Z(4430)} \downarrow \, , \,\,\,\,\,\,\lambda_{Z(4430)} \downarrow \, ,\nonumber\\
M_{Z_c(3900)}\downarrow  &\longmapsto&  M_{Z(4430)} \uparrow \, , \,\,\,\,\,\,\lambda_{Z(4430)} \uparrow \, ,\nonumber\\
\lambda_{Z_c(3900)}\uparrow  &\longmapsto&  M_{Z(4430)} \uparrow \, , \,\,\,\,\,\,\lambda_{Z(4430)} \uparrow \, ,\nonumber\\
\lambda_{Z_c(3900)}\downarrow  &\longmapsto&  M_{Z(4430)} \downarrow \, , \,\,\,\,\,\,\lambda_{Z(4430)} \downarrow \, ,
\end{eqnarray}
from the QCD sum rule in Eq.(11) at the energy scale $\mu=1.5\,\rm{GeV}$, the small variations  of the $M_{Z_c(3900)}$ and $\lambda_{Z_c(3900)}$
can lead to rather large changes of the $M_{Z(4430)}$ and $\lambda_{Z(4430)}$. In Eqs.(21-22), we take the experimental values $M_{Z_c(3900)}=3899\,\rm{MeV}$ and $M_{Z(4430)}=4475\,\rm{MeV}$ as input parameters, so the fitted parameters $\lambda_{Z_c(3900)}$ and 
$\lambda_{Z(4430)}$  are not as robust as the ones from the  QCD sum rules in Eqs.(15-19).

We may expect to calculate the masses and pole residues of the ground state and the first radial excited state of the $1^{+-}$ tetraquark states at
 the same energy scale. In Fig.4,  the masses of the ground state and the first radial excited state  are plotted   with variations of the  Borel parameters $T^2$ and energy scales $\mu$. From the figure, we can see that the masses decrease monotonously with increase of the energy scales, it is impossible to reproduce the experimental values $M_{Z_c(3900)}=(3899.0\pm 3.6\pm 4.9)\,\rm{ MeV}$ and $M_{Z(4430)}=(4475\pm7\,{_{-25}^{+15}})\,\rm {MeV}$ at the same energy scale, just as in the case of the
 $\rho(\rm 1S,2S)$, $\psi(\rm 1S,2S)$ and $\Upsilon(\rm 1S,2S)$ \cite{Baxi-G}.

 In this article, we take the threshold parameters as $\sqrt{s_0}=(4.7-4.9)\,\rm{GeV}$ and Borel parameters as $T^2=(2.7-3.3)\,\rm{GeV}^2$, then
  \begin{eqnarray}
  \exp\left(-\frac{s_0}{T^2} \right)&=&e^{-8.9}\sim e^{-6.7}\, ,
  \end{eqnarray}
  the continuum states are greatly depressed. The predictions are not sensitive to the continuum threshold parameters, although the masses and pole residues increase with increase of the threshold parameters. At the Borel window $T^2=(2.7-3.3)\,\rm{GeV}^2$, the masses and pole residues are rather stable with variations of the Borel parameters, platforms appear, so the predictions are reasonable.

\begin{figure}
 \centering
 \includegraphics[totalheight=8cm,width=12cm]{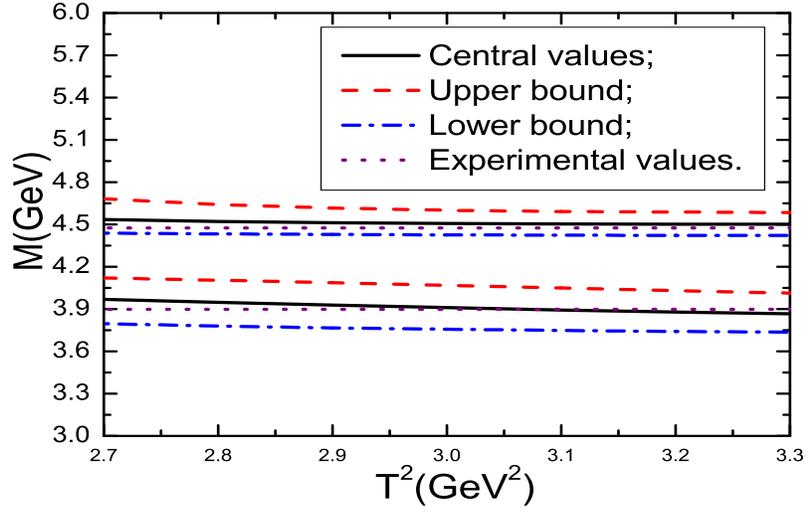}
    \caption{The masses of the ground state and the first radial excited state of the $1^{+-}$ tetraquark states with variations of the Borel parameters $T^2$. }
\end{figure}

\begin{figure}
 \centering
 \includegraphics[totalheight=8cm,width=12cm]{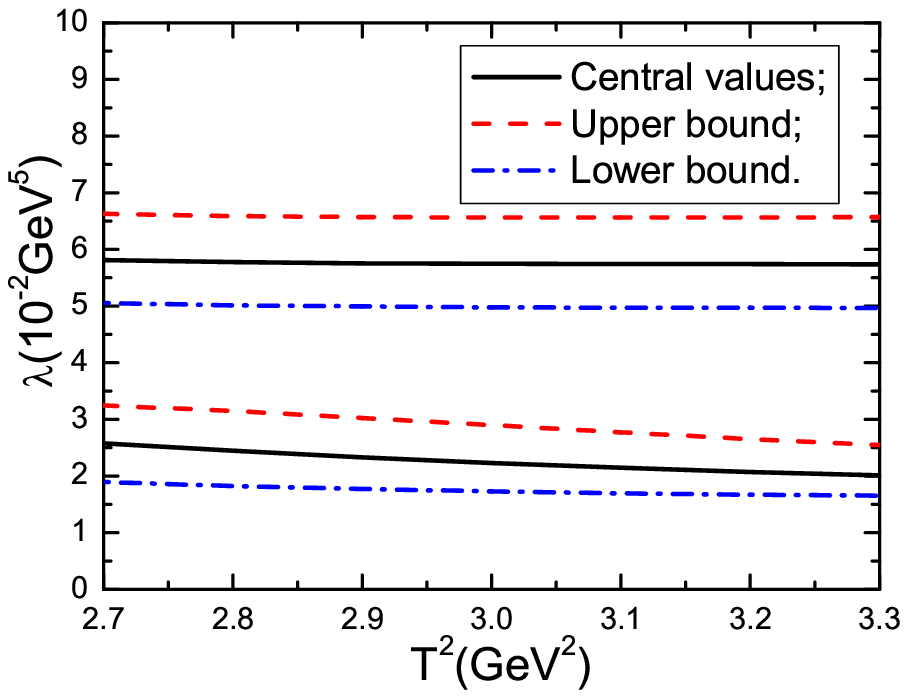}
    \caption{The pole residues of the ground state and the first radial excited state of the $1^{+-}$ tetraquark states with variations of the Borel parameters $T^2$. }
\end{figure}

\begin{figure}
 \centering
 \includegraphics[totalheight=8cm,width=12cm]{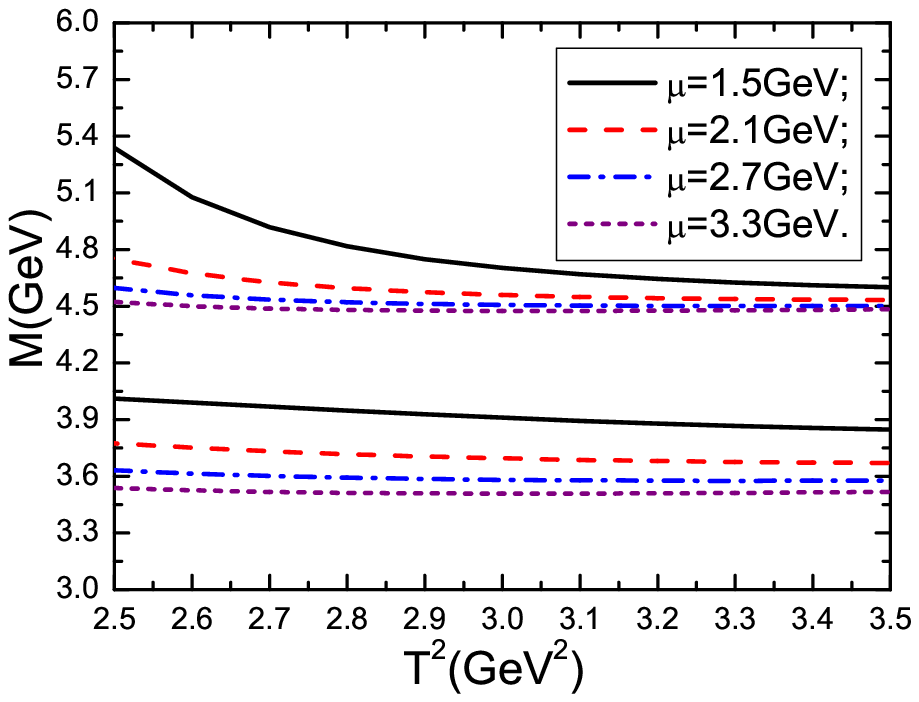}
    \caption{The masses of the ground state and the first radial excited state of the $1^{+-}$ tetraquark states with variations of the energy scales $\mu$ and Borel parameters $T^2$. }
\end{figure}

Now we perform Fierz re-arrangement to the current $J_\mu$  both in the color and Dirac-spinor  spaces and obtain the following result,
\begin{eqnarray}
J^{\mu} &=&\frac{1}{2\sqrt{2}}\left\{\,i\bar{c}i\gamma_5 c\,\bar{d}\gamma^\mu u-i\bar{c} \gamma^\mu c\,\bar{d}i\gamma_5 u+\bar{c} u\,\bar{d}\gamma^\mu\gamma_5 c-\bar{c} \gamma^\mu \gamma_5u\,\bar{d}c\right. \nonumber\\
&&\left. - i\bar{c}\gamma_\nu\gamma_5c\, \bar{d}\sigma^{\mu\nu}u+i\bar{c}\sigma^{\mu\nu}c\, \bar{d}\gamma_\nu\gamma_5u
- i \bar{c}\sigma^{\mu\nu}\gamma_5u\,\bar{d}\gamma_\nu c+i\bar{c}\gamma_\nu u\, \bar{d}\sigma^{\mu\nu}\gamma_5c   \,\right\} \, ,
\end{eqnarray}
the components such as $\bar{c}i\gamma_5 c\,\bar{d}\gamma^\mu u$, $\bar{c} \gamma^\mu c\,\bar{d}i\gamma_5 u$, etc couple   to the  meson-meson  pairs,
the strong decays
\begin{eqnarray}
Z^{\pm}_c(3900)(1^{+-}) &\to& h_c({\rm 1P})\pi^{\pm}\, , \, J/\psi\pi^{\pm}\, , \, \eta_c \rho^{\pm} \, , \, \eta_c(\pi\pi)_{\rm P}^{\pm} \, , \nonumber\\
Z^{\pm}(4430)(1^{+-}) &\to& h_c({\rm 2P})\pi^{\pm}\, , \, \psi^{\prime}\pi^{\pm}\, , \, \eta_c^{\prime} \rho^{\pm} \, , \, \eta_c^{\prime}(\pi\pi)_{\rm P}^{\pm} \, , \,h_c({\rm 1P})\pi^{\pm}\, , \, J/\psi\pi^{\pm}\, , \, \eta_c \rho^{\pm} \, , \, \nonumber\\
&&\eta_c(\pi\pi)_{\rm P}^{\pm} \, , \, \eta_c h_1(1170)^\pm\, , \, (D_0^*(2400)D)^\pm \, , \, (D^*D^*)^{\pm}\, ,
\end{eqnarray}
are Okubo-Zweig-Iizuka (OZI)  super-allowed, we take the decays to the $(\pi\pi)_{\rm P}^{\pm}$   final states as OZI super-allowed according to the decays $\rho \to \pi\pi$.
We can search for the $Z^{\pm}_c(3900)(1^{+-})$ and $Z^{\pm}(4430)(1^{+-})$ in those strong decays.

\section{Conclusion}
In this article, we  take the $Z_c(3900)$ and $Z(4430)$  as the ground state and the first radial excited state of the axial-vector  tetraquark states
 respectively with the symbolic quark structure
$[cu]_{S=1}[\bar{c}\bar{d}]_{S=0}-[cu]_{S=0}[\bar{c}\bar{d}]_{S=1}$,
and study their masses and pole residues with the QCD sum rules by  calculating the contributions of the vacuum condensates up to dimension-10 in a
consistent way in the operator product expansion.  The numerical result favors assigning  the $Z_c(3900)$ and $Z(4430)$
 as the ground state and the first radial excited axial-vector  tetraquark states, respectively. We can search for the $Z_c(3900)$ and $Z(4430)$ in the OZI  super-allowed
decays listed in Sect.3 in the future.
\section*{Acknowledgements}
This  work is supported by National Natural Science Foundation,
Grant Number 11375063, and Natural Science Foundation of Hebei province, Grant Number A2014502017.

\end{document}